\documentclass[twocolumn,showpacs,preprintnumbers]{revtex4}%
\usepackage{amsmath}
\usepackage{amsfonts}
\usepackage{amssymb}
\usepackage{graphicx}%
\providecommand{\U}[1]{\protect\rule{.1in}{.1in}}
\begin{document}
\title{A simple solvable energy landscape model that shows a thermodynamic phase
transition and a glass transition }
\author{Gerardo G. Naumis}
\affiliation{Instituto de F\'{\i}sica, Universidad Nacional Aut\'{o}noma de M\'{e}xico
(UNAM), Apartado Postal 20-364, 01000, M\'{e}xico, Distrito Federal, Mexico.}
\date{\today }

\pacs{*****}

\begin{abstract}
When a liquid melt is cooled,\ a glass or \ phase transition can be obtained
depending on the cooling rate. Yet, this behavior has not been clearly
captured in energy landscape models. Here a model is provided in which two key
ingredients are considered based in the landscape, metastable states and their
multiplicity. Metastable states are considered as in two level system models.
However, their multiplicity and topology allows a phase transition in the
thermodynamic limit, while a transition to the glass is obtained for fast
cooling. By solving the corresponding master equation, the minimal speed of
cooling required to produce the glass is obtained as a function of the
distribution of metastable and stable states. This allows to understand
cooling trends due to rigidity considerations in chalcogenide glasses.

\end{abstract}
\maketitle

\address{$^{1}$Instituto de Fisica, Universidad Nacional Aut\'{o}noma de
M\'{e}xico (UNAM)\\
Apartado Postal 20-364, 01000, Distrito Federal, Mexico.}

Humankind has been using glassy materials since the dawn of civilization.
However, their process of formation still poses many questions \cite{Anderson}%
\cite{Relaxation}\cite{Micoulaut}\cite{Kerner2}\cite{Mauro1}\cite{Mauro2}%
\cite{Magdaleno}. Glasses do not have long range order and are formed out of
thermal equilibrium, resulting in a limited use of the traditional tools of
the trade in solid state and statistical mechanics. Moreover, numerical
simulations are not able to provide definitive answers, since cooling speeds
achieved in numerical simulations are orders of magnitude higher than in real
cases \cite{Debenebook}. One of the main issues is the nature of the glass
transition \cite{Debenedetti}, for example, is it a purely dynamical effect or
there is a underlying thermodynamical singularity? The answer to this question
has practical implications, as how to calculate the minimal cooling speed
depending on the chemical composition in order to form a glass, or why some
chemical compounds form glasses while others will never reach such state
\cite{Phillips1}. Concerning this relationship between chemical composition
and minimal cooling speed, Phillips\cite{Phillips1} observed that for several
chalcogenides, this minimal speed is a function of rigidity. This initial
observation was the ingnition spark for the extensive investigation on
rigidity of glasses\cite{Thorpe0}\cite{Boolchand3}\cite{Georgiev}%
\cite{Naumis}\cite{Huerta0}\cite{Huerta1}\cite{Huerta}\cite{Huertaprb}, yet
this basic observation has not been quantitatively explained.

On the other hand, the energy landscape has been a useful picture to
understand glass transition \cite{Debenedetti}. However, due to its complicate
high dimensional topology, it is difficult to obtain closed analytical
results. It is not even clear how a phase transition is related with the
topology of the landscape, i.e., why a global minimum leads to singularities
in the thermodynamical behavior. Clearly, there is a lack of a minimal simple
solvable model of landscape that can display a phase and a glass transition
depending on the cooling rate. Here we present such model by combining the two
most basic ingredients that are belived to be fundamental in the problem.
Furthermore, the model allows to get a glimpse on the connection between
minimal cooling speed, energy landscapes, rigidity, and Boolchand intermediate
phases \cite{IBoolchand1}\cite{IBoolchand3}.

\begin{figure}
[ptb]
\begin{center}
\includegraphics[
height=2.023in,
width=2.2444in
]%
{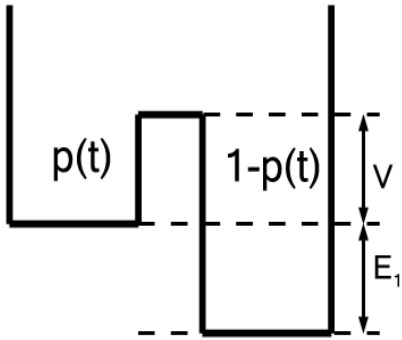}%
\caption{The two level system energy landscape, showing the barrier height $V$
and the asymmetry $E_{1}$ between the two levels. The population of the upper
well is $p(t)$.}%
\label{figurewells}%
\end{center}
\end{figure}

The first ingredient is based in a well known fact: glasses are trapped in
metastable states, while crystals are global minimums in the landscape. A
common way to describe the corresponding physics is through the use of two
level system (TLS). If the glassy metastable state has energy $E_{1}$ and the
crystalline global minimum an energy $E_{0}=0$, the system is trapped in the
glassy state due to an energy barrier $V$ measured from $E_{1}$, as seen in
Fig. \ref{figurewells}. Following Huse et. al.\cite{Huse} and Langer et. al.
\cite{Langerprl}\cite{Langer}\cite{Langer1990}, who described the residual
population of the metastable state for a TLS at zero temperature, the cooling
process can be described by a master equation, in which the probabibilty
$p(t)$ of finding the system in the metastable state, assuming that the system
is in contact with a bath at temperature $T$, is\cite{Langerprl},
\begin{equation}
\frac{dp(t)}{dt}=-\Gamma_{\uparrow\downarrow}p(t)+\Gamma_{\downarrow\uparrow
}\left[  1-p(t)\right]  , \label{master0}%
\end{equation}
where $\Gamma_{\uparrow\downarrow}$ is the transition rate from the upper well
to the lower, and the transitions from the lower to the upper take place at
rate $\Gamma_{\downarrow\uparrow}=e^{-\frac{E_{1}}{T}}\Gamma_{\uparrow
\downarrow}$. If quantum mechanical tunneling is neglected, $\Gamma
_{\uparrow\downarrow}=$ $\Gamma_{0}e^{-\frac{V}{T}}$, where $\Gamma_{0}$ is a
small frequency of oscillation at the bottom of the walls.%

%TCIMACRO{\FRAME{ftbpFU}{2.2444in}{2.023in}{0pt}{\Qcb{The two level system
%energy landscape, showing the barrier height $V$ and the asymmetry $E_{1}$
%between the two levels. The population of the upper well is $p(t)$.}%
%}{\Qlb{figurewells}}{fig1relaxation.eps}%
%{\special{ language "Scientific Word";  type "GRAPHIC";  display "USEDEF";
%valid_file "F";  width 2.2444in;  height 2.023in;  depth 0pt;
%original-width 8.2607in;  original-height 11.9312in;  cropleft "0";
%filename 'fig1Relaxation.eps';file-properties "XNPEU";}} }%
%BeginExpansion
%EndExpansion

Eq. (\ref{master0}) describes the relaxation towards $p_{0}(T)$, the
population at thermal equilibrium obtained from the stationary condition, as
can be seen by rewriting Eq. (\ref{master0}) as \cite{Langerprl},%

\begin{equation}
\frac{dp(t)}{dt}=\Gamma_{\uparrow\downarrow}(1+e^{-\frac{E_{1}}{T}}%
)(p_{0}(T)-p(t)), \label{masterelax}%
\end{equation}
where $p_{0}(T)$ is given by,%
\begin{equation}
p_{0}(T)=\frac{e^{-\frac{E_{1}}{T}}}{1+e^{-\frac{E_{1}}{T}}}.
\end{equation}
When the system is cooled by a given protocol $T=T(t)$, it can be proved that
at zero temperature there is a probability $p(T=0)$ for the system to be in
the metastable state, which is indicative of a glassy behavior \cite{Langer}%
\cite{Brey}. This simple model is very appealing and can be used to explain
low temperature anomalies in glasses\cite{WAPhillips}\cite{Anderson2}.
However, a huge part of the physics is missing: the system does not present a
phase transition at low cooling speeds. To achive this goal, here we introduce
a key element to the TLS landscape topology: the multiplicity of states.
Again, there is a common agreement that the number of metastable states is
much bigger than their crystalline counteparts. Assume that the energy $E_{1}$
has a degeneracy $g_{1}$, while the ground energy $E_{0}$ has degeneracy
$g_{0}$, thus Eq. (\ref{master0}) needs to be modified to take into account
transitions between different states that are in the low and upper wells. Call
$p_{\uparrow s}(t)$ the population of one of these\ $g_{1}$ in the upper
states, and $p_{\downarrow s}(t)$ the population of one of these\ $g_{0}$ in
the low states. Eq. (\ref{master0}) becomes,%

\begin{align}
\frac{dp_{\uparrow s}(t)}{dt}  &  =-\sum_{l\neq s}^{g_{1}-1}\Gamma
_{\uparrow\uparrow}^{sl}p_{\uparrow s}(t)-\sum_{m}^{g_{0}}\Gamma
_{\uparrow\downarrow}^{sm}p_{\uparrow s}(t)\\
&  +\sum_{l\neq s}^{g_{1}}\Gamma_{\uparrow\uparrow}^{ls}p_{\uparrow l}%
(t)+\sum_{m}^{g_{0}}\Gamma_{\downarrow\uparrow}^{ms}p_{\downarrow m}(t),
\end{align}
where $\Gamma_{\uparrow\uparrow}^{sl}$ denotes the transition rate from state
$s$ to $l$, both in the upper well. The notation for the other transition
rates is similar, and an equivalent expression can be written for
$dp_{\downarrow s}(t)/dt$. To formulate the model, we use the simplest
topology, i.e., all metstable states are connected within them with the same
transition rate, i.e., $\Gamma_{\uparrow\uparrow}^{sl}\equiv\Gamma
_{\uparrow\uparrow}$. A similar situation holds for the crystalline states
$\Gamma_{\downarrow\downarrow}^{sl}\equiv\Gamma_{\downarrow\downarrow}$.
Transitions between up and lower states have also the same probability
$\Gamma_{\uparrow\downarrow}^{sl}\equiv\Gamma_{\uparrow\downarrow}$ and
$\Gamma_{\downarrow\uparrow}^{sl}=\Gamma_{\downarrow\uparrow}$. Under such
simple landscape topology, the previous master equation can be reduced to,%
\begin{equation}
\frac{dp(t)}{dt}=-g_{0}\Gamma_{\uparrow\downarrow}p(t)+g_{1}\Gamma
_{\downarrow\uparrow}\left[  1-p(t)\right]  \label{masterdege}%
\end{equation}
where now $p(t)=%
%TCIMACRO{\dsum \limits_{s=1}^{g_{1}}}%
%BeginExpansion
{\displaystyle\sum\limits_{s=1}^{g_{1}}}
%EndExpansion
p_{\uparrow s}(t)$ is the total probability of finding the system in the upper
well. Let us show how Eq. (\ref{masterdege}) can give a phase transition under
thermal equilibrium conditions. In that case, $dp(t)/dt=0$ and,%
\begin{equation}
p_{0}(T)=\frac{g_{1}\Gamma_{\downarrow\uparrow}}{g_{0}\Gamma_{\uparrow
\downarrow}+g_{1}\Gamma_{\downarrow\uparrow}}=\frac{(g_{1}/g_{0}%
)e^{-\frac{E_{1}}{T}}}{1+(g_{1}/g_{0})e^{-\frac{E_{1}}{T}}}.
\end{equation}

A phase transition can occur if $(g_{1}/g_{0})e^{-\frac{E_{1}}{T}}$ becomes
discontinous in the thermodinamical limit. The most simple example is the
following. Suppose that we have $N$ particles, and the potential is such that
the crystalline state is unique ($g_{0}=1$), with energy $E_{0}=0$, and assume
that the number of metastable states grows exponentially with $N$, as is the
case in many glassy systems \cite{Debenebook} where $g_{1}=e^{N\ln\Omega
(E_{1})}$. $\Omega(E_{1})$ is a measure of the landscape complexity
\cite{Debenedetti}. Also, the only way to make $\left\langle E\right\rangle $
an intensive quantity with only one energy is to have $E_{1}=N\epsilon$, where
$\epsilon$ is an energy per particle. As an example, this behavior can be
readly obtained when two particles, confined in cells, interact with
neighboring cells as in nearly one dimensional models of magnetic walls
\cite{Chandler}. For this particular case, $g_{1}=2^{N}$ and $g_{0}=1$. Using
the previous general considerations, $p_{0}(T)$ can be written as,%
\begin{equation}
p_{0}(T)=\frac{e^{\left(  \ln\Omega(E_{1})-\frac{\epsilon}{T}\right)  N}%
}{1+e^{\left(  \ln\Omega(E_{1})-\frac{\epsilon}{T}\right)  N}}=\frac{z^{N}%
}{1+z^{N}} \label{equilibrium}%
\end{equation}
with $z=[\exp(\Omega(E_{1})-\epsilon/T)N]$. In the thermodynamic limit
\ $N\rightarrow\infty$, the function $f(z)=z^{N}$ develops a discontinuty at
$z=1$, and it is easy to see that there is a phase transition at temperature,
\begin{equation}
T_{c}=\frac{\epsilon}{\ln\Omega(E_{1})}%
\end{equation}
with a discontinuous specific heat,%
\begin{equation}
c\equiv\epsilon\frac{dp_{0}(T)}{dT}=\left\{
\begin{array}
[c]{c}%
0\text{ if }T\neq T_{c}\\
\infty\text{ if }T=T_{c}%
\end{array}
\right.  \label{specific}%
\end{equation}
Now the model is able to produce a phase transition under thermal equilibrium.
This can be clearly seen in Fig. 2 for $g_{1}=2^{N}$ and $g_{0}=1$, where we
plot Eq. (\ref{equilibrium}) for different values of $N$ using dotted lines.
Notice how the phase transition is built by a progressive sharpening of the
jump in $p_{0}(T)$ as $N$ grows. According to Eq. (\ref{specific}), the
specific heat is just the derivative of $p_{0}(T)$, thus the sharpening leads
to the singularity in the thermodynamical limit.

Now we will show that a glassy behavior is obtained for fast enough cooling.
To solve Eq. (\ref{masterdege}), one needs to specify the cooling protocol
$T=T(t)$, and write the master equation in terms of a dimensionless cooling
rate. Two kinds of protocols are useful \cite{Langerprl}\cite{Langer}, one is
the linear cooling $T=T_{0}-rt$, used mainly in experiments, and the
hyperbolic one $T=T_{0}/(1+Rt)$, which allows a simple handling of the
asymptotics involved. For the hyperbolic case, the master equation can be
written as,%
\begin{equation}
\delta\frac{dp(x)}{dx}=-g_{1}x^{\mu}+(g_{0}+g_{1}x^{\mu})p(x)
\label{linearcooling}%
\end{equation}
where $x=\exp(-V/T)$ and $\delta=RV/\Gamma_{0}T_{0}$. The parameter $\mu
=E_{1}/V$ measures the asymmetry of the well. The linear case also follows Eq.
(\ref{linearcooling}), since one can rescale the boundary layer \cite{Langer}
that appears in Eq. (\ref{linearcooling}), leading to the same hyperbolic
equation with $\delta=rV/T_{0}$. Eq. (\ref{linearcooling}) can be solved to
give,
\begin{align}
p(x)  &  =\exp\left[  \frac{1}{\delta}\left(  g_{0}x+\frac{g_{1}x^{1+\mu}%
}{1+\mu}\right)  \right]  \times\\
&  \left\{  p(0)-\frac{g_{1}}{\delta}\int_{0}^{x}y^{\mu}\exp\left[  -\frac
{1}{\delta}\left(  g_{0}y+\frac{g_{1}y^{1+\mu}}{1+\mu}\right)  \right]
\right\}  .\nonumber
\end{align}
As an example, Fig. 2 shows $p(x)$ for different cooling rates and system
sizes, using a linear cooling and $g_{1}=2^{N}$, $g_{0}=1$, compared with the
equilibrium distribution that develops a phase transition at $T_{c}$. Notice
in Fig. 2 that $p(0)$ is the residual population at $T=0$, indicative of a
glassy behavior. Also, the slope of $dp(T)/dT$ does not tend to infinity, and
the corresponding specific heat $c$ in no longer discontinuous, as in real
glass transitions.

We can obtain the analytical value of $p(0)$ by assuming that the system was
at thermal equilibrium before being cooled at a temperature $T_{0}>>T_{c}$. In
that case $x\rightarrow1$, and the population is given by the equilibrium
distribution, $p_{0}(x_{0})$ $=(g_{1}/g_{0})x_{0}^{\mu}/((g_{1}/g_{0}%
)x_{0}^{\mu}+1)$ where $x_{0}=\exp(-V/T_{0})$ . From Eq. (\ref{linearcooling}%
), we obtain a general expression for $p(0)$,%
\begin{align}
p(0) &  =\frac{x_{0}^{\mu}}{x_{0}^{\mu}+(g_{0}/g_{1})}\exp\left[  -\frac
{1}{\delta}\left(  g_{0}x_{0}+\frac{g_{1}x_{0}^{1+\mu}}{1+\mu}\right)
\right]  \label{p0}\\
&  +\frac{g_{1}}{\delta}\int_{0}^{x_{0}}y^{\mu}\exp\left[  -\frac{1}{\delta
}\left(  g_{0}y+\frac{g_{1}y^{1+\mu}}{1+\mu}\right)  \right]  dy.\nonumber
\end{align}
Zero population is only achieved if both terms in Eq. (\ref{p0part}) are zero,
as is the case for $\delta\rightarrow0$. Then we recover the phase transition,
a fact that makes us confident in the result. To understand more deeply Eq.
(\ref{p0}), let us study the particular case $g_{1}=e^{N\ln\Omega(E_{1})}$ and
$g_{0}=1$, with $E_{1}=N\epsilon$. The second integral contains the term
$g_{1}y^{1+\mu}\approx\exp\left[  (\ln\Omega(E_{1})-\varepsilon/T)N\right]  $,
which can be $0$ or $\infty$ in the thermodynamical limit depending whether
$y<x_{c}$ or $y\geq x_{c}$, where,%
\begin{equation}
x_{c}\equiv\exp(-V/T_{c})=\Omega(E_{1})^{-V/\epsilon}.
\end{equation}
If $V$ does not scale with $N$, for big $N$ Eq. (\ref{p0}) can be written as,%
\begin{equation}
p(0)\approx\frac{x_{0}^{\mu}}{x_{0}^{\mu}+(g_{0}/g_{1})}\exp\left[
-\frac{g_{1}x_{0}^{\mu}}{\delta\mu}\right]  +g_{1}\gamma(1+\mu,x_{c}%
/\delta)\delta^{\mu},\label{p0part}%
\end{equation}
here $\gamma$ is the lower incomplete gamma function. The evolution of the
residual population given by Eq. (\ref{p0part}) is shown in Fig.
(\ref{residualspeed}) as a function of the cooling speed and system size.
%TCIMACRO{\FRAME{ftbpFU}{2.6327in}{2.4076in}{0pt}{\Qcb{Population as a function
%of the temperature using a linear cooling for different number of particles
%$N=2,4$ and $8$, with $V=1.0$, $\varepsilon=1$, $R=1.4$ and $T_{0}/T_{c}=72$ ,
%$g_{1}=2^{N}$, $g_{0}=1$ obtained by solving the master equation. The
%equilibrium population, obtained for $\delta\rightarrow0$ is also shown, as
%indicated in the inset. Notice how the phase transition is built by a
%progressive sharpening of the jump in $p(T)$ as $N$ grows.}}{}{population.eps}%
%width 2.6327in;  height 2.4076in;  depth 0pt;  original-width 7.9554in;
%original-height 7.2709in;  cropleft "0";  croptop "1";  cropright "1";
%cropbottom "0";  filename 'population.eps';file-properties "XNPEU";}} }%
%BeginExpansion
\begin{figure}
[ptb]
\begin{center}
\includegraphics[
height=2.4076in,
width=2.6327in
]%
{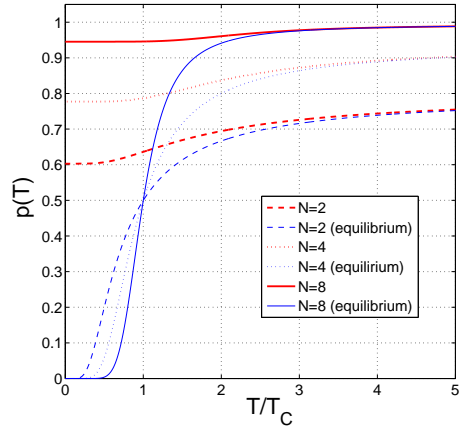}%
\caption{Population as a function of the temperature using a linear cooling
for different number of particles $N=2,4$ and $8$, with $V=1.0$,
$\varepsilon=1$, $R=1.4$ and $T_{0}/T_{c}=72$ , $g_{1}=2^{N}$, $g_{0}=1$
obtained by solving the master equation. The equilibrium population, obtained
for $\delta\rightarrow0$ is also shown, as indicated in the inset. Notice how
the phase transition is built by a progressive sharpening of the jump in
$p(T)$ as $N$ grows.}%
\end{center}
\end{figure}
%EndExpansion
As a general trend, the cooling speed required to make a glass with fixed
$p(0)$ increases with the system size. Also, it is possible to observe \ a
crossover which separates different behaviors of $p(0)$. For example, in Fig.
(\ref{residualspeed}), if $N=4,5,6$ and $7$, $p(0)$ begins to increase for a
high $\delta$ after it reaches a plateau that begins around $\delta\approx1$.
The same increase is observed for $N=9$ and $10$, although shifted to the
right in such a way that it does not appear in the current plot. \ This
crossover is due to the different growing speeds in Eq. (\ref{p0part}). The
first term of Eq. (\ref{p0part}) goes to zero if,%
\begin{subequations}
\begin{equation}
\delta<<\Omega(E_{1})^{N}\frac{V}{E_{1}}e^{-^{E_{1}/T_{0}}}\equiv\delta
_{c}(N),\label{Bound1}%
\end{equation}
which defines a critical speed $\delta_{c}(N)$. For $\delta>\delta_{c}(N)$,
$p(0)$ is dominated by the first term in Eq. (\ref{p0part}). The remaining
term in Eq. (\ref{p0part}) regulates the residual population for lower speeds
$\delta<\delta_{c}(N)$. This term produces the plateau at a saturating value
of $p(0)$,%
\end{subequations}
\begin{equation}
p(0)\approx\frac{1}{1+(N\epsilon/V)}\equiv p_{s}(0).
\end{equation}
For a finite $N$, this implies that there are two kinds of glassy phases, one
obtained for intermediate cooling rates in which $p(0)$ reaches a limiting
value. The other kind is obtained for $\delta>\delta_{c}(N)$.
%TCIMACRO{\FRAME{ftbpFU}{2.4408in}{2.1982in}{0pt}{\Qcb{Residual population at
%$T=0$ as a function of the cooling speed for different number of particles
%$N=10,15,20,25$ and $30$, with $V=0.5$, $\varepsilon=1$, $R=1.4$, $g_{1}%
%=2^{N}$, $g_{0}=1$ and $T_{0}/T_{c}=72$.}}{\Qlb{residualspeed}}%
%{residualpopulation2.eps}{\special{ language "Scientific Word";
%type "GRAPHIC";  maintain-aspect-ratio TRUE;  display "USEDEF";
%valid_file "F";  width 2.4408in;  height 2.1982in;  depth 0pt;
%original-width 5.9591in;  original-height 5.3632in;  cropleft "0";
%filename 'residualpopulation2.eps';file-properties "XNPEU";}} }%
%BeginExpansion
\begin{figure}
[ptb]
\begin{center}
\includegraphics[
height=2.1982in,
width=2.4408in
]%
{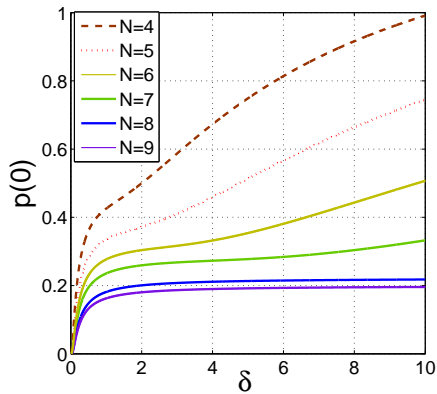}%
\caption{Residual population at $T=0$ as a function of the cooling speed for
different number of particles $N=10,15,20,25$ and $30$, with $V=0.5$,
$\varepsilon=1$, $R=1.4$, $g_{1}=2^{N}$, $g_{0}=1$ and $T_{0}/T_{c}=72$.}%
\label{residualspeed}%
\end{center}
\end{figure}
%EndExpansion

However, since $p_{s}(0)$ goes to zero as $N\rightarrow\infty$, $\delta
_{c}(N)$ turns out to be the minimal speed required to make a glass in the
thermodynamical limit, and leads to a critical $R_{c}$,%
\begin{equation}
R_{c}=\Omega(E_{1})^{N}\left(  \frac{T_{0}}{E_{1}}\right)  e^{-^{E_{1}/T_{0}}%
}\Gamma_{0}.
\end{equation}
From an analysis of Eq. (\ref{masterelax}) and Eq. (\ref{master0}), it is easy
to see that $R_{c}$ is basically the inverse relaxation time for
crystalization. Although it is surprising that $R_{c}$ does not depend on $V$,
this is a result of the assumption that $V$ does not scale with $N$. If this
is the case, the term $g_{1}x_{0}^{1+\mu}$ in Eq. (\ref{p0}) can determine if
$g_{0}x_{0}$ plays a role in the first exponential, leading to a critical
speed that depends on $V$. Notice that the result is in agreement with the
remarkable observation made by Phillips concerning chemical composition and
minimal cooling speed required to make glasses\cite{Phillips1}, since $R_{c}$
depends on the number of metastable states. Rigidity provides such an indirect
count of metastable and stable states \cite{NaumisLandscape}%
\cite{NaumisLandscape2}.

\ In conclusion, we have introduced the topology of the energy landscape in a
two level model of glass. As a result, we have a solvable model that has a
thermodynamic phase transition for low cooling rates and a glass transition
for fast cooling.

\textbf{Acknowledgments.} I would like to thank Denis Boyer for useful
suggestions and a critical reading of the manuscript. This work was supported
by DGAPA UNAM project IN100310-3. Calculations were made at Kanbalam
supercomputer at DGSCA-UNAM.

\end{document}